\documentstyle[twocolumn,epsf]{jpsj}

\def\runtitle{ Anomalous X-ray Scattering}
\def\runauthor{Manabu {\sc Takahashi}, Jun-ichi {\sc Igarashi}
and Peter {\sc Fulde}}
\title{Pre-K-Edge Structure on Anomalous X-Ray Scattering in LaMnO$_3$}

\author{Manabu {\sc Takahashi}$^1$, Jun-ichi {\sc Igarashi}$^1$
and Peter {\sc Fulde}$^2$}
\inst{
$^1$Faculty of Engineering, Gunma University, Kiryu, Gunma 376-8515\\
$^2$Max-Planck-Institut f\"ur Physik komplexer Systeme,\\
N\"othnitzer Str. 38, D-01187 Dresden, Germany}

\recdate
{\ \ \ \ \ \ \ \ \ \ \ \ \ \ \ \ \ \ \ \
}

\abst{
We study the pre-K-edge structure of the resonant X-ray scattering 
for forbidden reflections (anomalous scattering) in LaMnO$_3$, 
using the band calculation based on the local density approximation. 
We find a two-peak structure with an intensity approximately $1/100$
of that of the main peak. This originates from
a mixing of $4p$ states of Mn to $3d$ states of neighboring Mn sites. 
The effect is enhanced by an interference with the tail of the main peak.
The effect of the quadrupole transition is found to be
one order of magnitude smaller than that of the dipole transition,
modifying slightly the azimuthal-angle dependence.
}
\kword
{resonant X-ray scattering, LaMnO$_3$, pre-peak of the K-edge,
quadrupole transition, cooperative Jahn-Teller effect, orbital ordering}
 
\begin{document}
\sloppy
\maketitle

Resonant X-ray scattering technique has recently attracted much
interest as a method of observing orbital order.
Murakami {\em et al.}\cite{Murakami} carried out such an experiment
with photon energy around  the K-edge of Mn in LaMnO$_3$, and
found that the forbidden $(h,0,0)$ reflection with $h$ being an odd integer
gains some intensity below 780K, where the crystal starts 
to distort.\cite{Murakami}
We call this phenomenon anomalous X-ray scattering (AXS).
The relevant states are $4p$ states of Mn.
They are highly extended in the crystal.
Therefore this AXS intensity does not directly reflect orbital order.
Ishihara and Maekawa\cite{Ishihara} suggested 
that the energy levels of $4p_x$ and $4p_y$ states are split by the 
anisotropic part of the intraatomic Coulomb interaction between $4p$ 
and $3d$ states and that its dependence on 
occupied orbitals $d_{3x^2-r^2}$ or $d_{3y^2-r^2}$ causes the AXS intensities. 
However, Elfimov {\em et al.},\cite{Elfimov} Benfatto {\em et al.}
\cite{Benfatto} and the present authors\cite{Taka1} have recently carried out
{\em ab-initio} calculations, and have explicitly shown that
neighboring oxygens strongly influence the $4p$ states of Mn through
a Jahn-Teller (JT) distortion, resulting in a large AXS intensity.
Those authors have also shown that the intraatomic Coulomb interaction has 
a small effect on the AXS intensity only.\cite{Elfimov,Benfatto,Taka1}

The above studies refer to AXS near the main peak.
Recently, pre-K-edge (PKE) structures have been observed in
the AXS intensity in other materials such as V$_2$O$_3$\cite{Paolasini} 
and in resonant magnetic X-ray scattering in CoO.\cite{Neubeck}
In this paper we focus our attention on the {\em PKE } structure of
the AXS intensities in LaMnO$_3$.
Although it has not been observed yet, 
we hope that this study encourages experimentalists to investigate 
the PKE structures in that material.

On each Mn atom in LaMnO$_3$, the doubly-degenerate $e_g$ orbitals 
are occupied by one electron only [$(t_{2g})^3(e_g)^1$ configuration
because of the cubic symmetry]. 
Spins of the $3d$ electrons are parallel on each atom due to a strong 
Hund-rule coupling, and show antiferromagnetic
long-range order.\cite{Wollan}
The orbital degeneracy is lifted by a JT distortion,
which takes place cooperatively in a crystal,\cite{Kanamori} 
or by the orbital exchange interaction, which is similar to the 
superexchange for spins.\cite{Kugel}
Independent of the particular driving force,
a sizable crystal distortion is realized.
In the actual crystal, the structure is more complicated because 
each MnO$_6$ octahedra is slightly tilted, implying
a $P_{nma}$ structure.\cite{Elemans}
For the crystal distortion of LaMnO$_3$, see Fig. 1 in Ref.~5.

\begin{figure}
\begin{center}
\epsfile{file=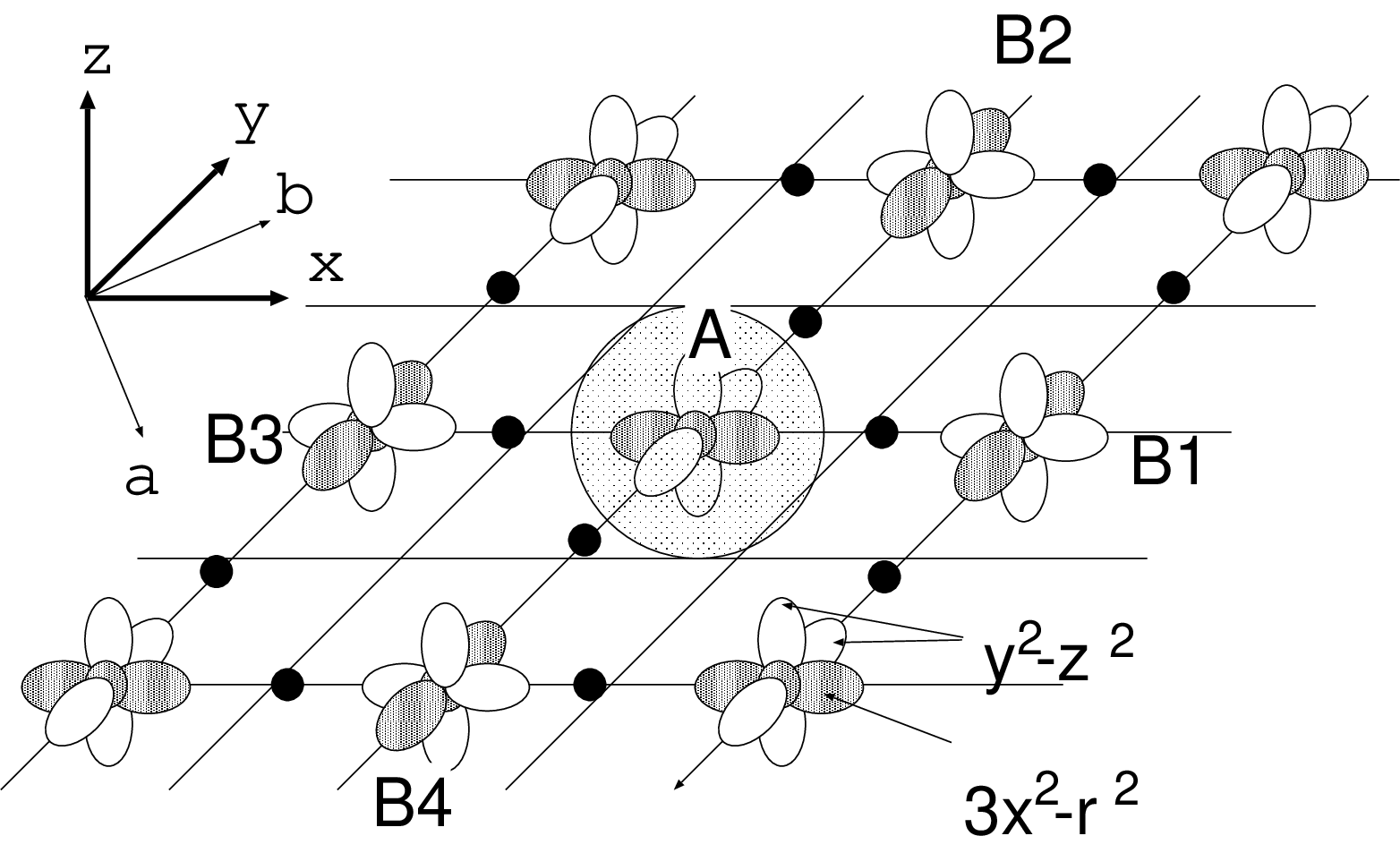,width=8cm}
\caption{Occupied (hatched lobes) and unoccupied (open lobes) $e_g$ 
orbitals for the majority spin states. 
Orbitals are ``antiferromagnetically" ordered in the $a-b$ plane
and ``ferromagnetically" ordered along the $c$ axis.
Solid circles represent oxygen atoms.
}\label{fig.system}
\end{center}
\end{figure}

We carry out a band calculation based on the local-density 
approximation (LDA) and the use of the KKR method,
just the same way as in Ref.~5, 
assuming the actual crystal distortion and tilt. 
As a result, we obtain an insulating ground state with a gap of 
$\sim 0.2$ eV (experimentally $\sim 0.24$ eV\cite{mahendiran}).
Spins align ferromagnetically in the $a-b$ plane and antiferromagnetically
along the $c$ axis (A-type antiferromagnet), with a magnetic moment of $3.35$ 
$\mu_{\rm B}$ per Mn atom.
Orbitals are ordered ``antiferromagnetically" in the $a-b$ plane
and ``ferromagnetically" along the $c$ axis.
These findings are consistent with experiments.
In case of orbital order, Mn sites are divided into two sublattices 
A and B such that at A sites the $d_{3x^2-r^2}$ state has
a larger occupation number than the $d_{y^2-z^2}$ state.
Figure \ref{fig.system} shows schematically this situation.
The present calculation gives for the occupation numbers of 
the $d_{3x^2-r^2}$ and $d_{y^2-z^2}$ orbitals the values $0.92$ and $0.57$, 
respectively, in fair agreement with previous calculations.\cite{Pickett}

\begin{figure}
\begin{center}
\epsfile{file=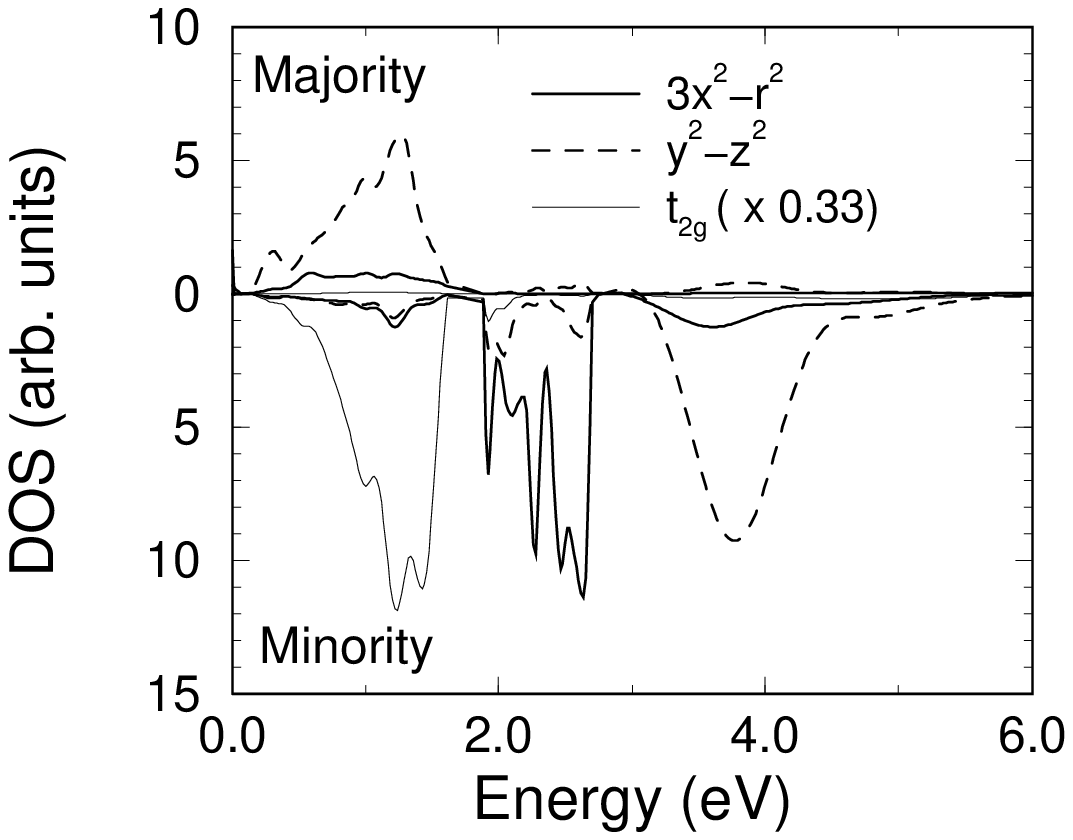,width=8cm}
\caption{Density of states projected onto the $d$ symmetry at A sites. 
The zero of energy is at the top of the valence band.
The intensity of the $t_{2g}$ symmetry is reduced to one-third of
its actual value.
}\label{fig.d}
\end{center}
\end{figure}

Figure \ref{fig.d} shows the density of states 
(DOS) projected onto states of $d$ symmetry above the top of the valence band
at A sites.
For the majority spin states, the $d_{y^2-z^2}$ DOS is dominant.
For the minority spin states, the $d_{t_{2g}}$, $d_{3x^2-r^2}$ and
$d_{y^2-z^2}$ DOS's appear separately with energies from 0 to 4 eV. 
These peaks are slightly different from the LDA$+U$ result,\cite{Elfimov} 
in which the partial DOS's of the majority spin states are 
overlapping with energies in the range of 2 to 6 eV. The LDA results
are expected to be better, since it is known from other oxides\cite{Taka2} 
that the $3d$ DOS in Hartree-Fock approximation is modified by electron
correlations to become close to that of the LDA.
 
\begin{figure}
\begin{center}
\epsfile{file=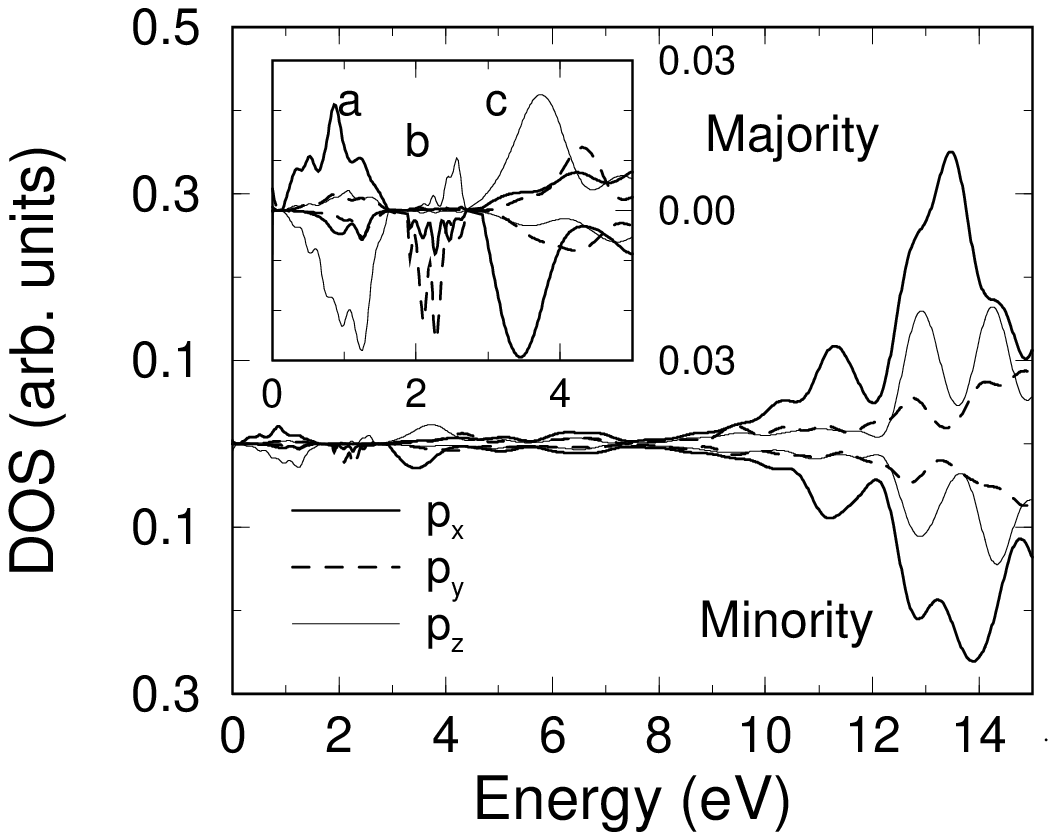,width=8cm}
\caption{Density of states projected onto the $p$ symmetry at A sites. 
The zero of energy is at the top of the valence band.
The inset shows the intensity in the pre-edge region.
}\label{fig.p}
\end{center}
\end{figure}

Figure \ref{fig.p} shows the DOS at A sites projected onto states of
$p$ symmetry.
The PKE structure appears at energies between $0\sim 5$ eV with
the intensity being roughly $1/20$ of the main peak intensity.
It originates from the $4p$ states mixing to the $3d$ states of 
neighboring Mn sites. As shown in the inset,
it is classified into three regions $a$, $b$ and $c$.
For the majority spin states, only the $p_x$ DOS has an appreciable 
intensity in region $a$, 
while the $p_z$ DOS is dominating for the minority spin states. 
The former results from the mixing to the unoccupied 
$d_{z^2-x^2}$ states of Mn sites in the same $a-b$ plane ($B_1$ and $B_3$ in 
Fig.~\ref{fig.system}), while the latter (minority spin) 
comes from the mixing 
to the $d_{y^2-z^2}$ states (majority spin) of Mn sites above and below 
the considered plane (note that the magnetization direction is opposite
in adjacent planes).
In region $b$, the $p_y$ DOS is most prominent for the minority spin states. 
It results from the mixing to 
the $d_{3y^2-r^2}$ state of Mn sites in the same $a-b$ 
plane ($B_2$ and $B_4$ in Fig.~\ref{fig.system}). 
In region $c$, the $p_z$ DOS is dominant for the majority spin states, 
while the $p_x$ DOS is dominant for the minority spin states. 
The former results from the mixing to the $d_{y^2-z^2}$ state 
(minority spin) of Mn sites above and below the considered plane, 
while the latter results
form the mixing to the $d_{z^2-x^2}$ (minority spin) of Mn site 
in the same $a-b$ plane (again $B_1$ and $B_3$ in Fig.~\ref{fig.system}).
This PKE structure is consistent with the previous LDA$+U$ result,
\cite{Elfimov} although the details are different.
 
The resonant X-ray scattering is described by a second-order process 
in which the photon is virtually absorbed and then emitted.
The dipole transition corresponds to a transition from the $1s$ state
to the $p$-symmetric states at Mn sites, while
the quadrupole transition corresponds to a transition from the $1s$
state to the $d$-symmetric states.
Since in the PKE region the $d$-symmetric DOS 
is much larger than the $p$-symmetric DOS,
the quadrupole effect can be comparable in size with the dipole effect.
In the followings we consider both possibilities.
The scattering geometry is the same as the experimental configuration
as regards the polarization, the Bragg angle $\theta$ and the azimuthal angle 
$\psi$.\cite{Murakami}
The condition $\psi=0$ corresponds to the configuration 
in which the $c$ axis is perpendicular to the diffraction plane.
We consider only the scattering from polarization $\sigma$ to $\pi'$,
since the $\sigma\to\sigma'$ scattering has no AXS intensity.\cite{com1}

The scattering amplitude $T_{\sigma\to\pi'}(\omega)$ at Mn sites 
is divided into dipole and quadrupole processes:
\begin{equation}
 T_{\sigma\to\pi'}(\omega) =J_{\sigma\to\pi'}(\omega)
                          +L_{\sigma\to\pi'}(\omega),
\end{equation}
where
\begin{equation}
  J_{\sigma\to\pi'}(\omega)
  = \sum_{\alpha\beta}(D^{\pi'})^\dagger_{\alpha}
    M_{\alpha\beta}(\omega)D^{\sigma}_{\beta},
\end{equation}
with
\begin{eqnarray}
 M_{\alpha\beta}(\omega) &=& \sum_j
  \frac{\langle 1s|x_\alpha|p_j\rangle\langle p_j|x_\beta|1s\rangle}
       {\omega-(\epsilon_j-\epsilon_{1s})+i\Gamma}, \\
 D^\sigma&=& \left( \begin{array}{c}
                 -\sin\psi/\sqrt{2}\\
                  \sin\psi/\sqrt{2}\\
                  \cos\psi
                  \end{array} \right), \\ %\quad
 D^{\pi'}&=& \left( \begin{array}{c}
                (\cos\theta-\sin\theta\cos\psi)/\sqrt{2}\\
                (\cos\theta+\sin\theta\cos\psi)/\sqrt{2}\\
                 -\sin\theta\sin\psi
                  \end{array} \right), 
\end{eqnarray}
and
\begin{equation}
 L_{\sigma\to\pi'}(\omega)
  = \sum_{\alpha\beta}(Q^{\pi'})^\dagger_{\alpha}
    N_{\alpha\beta}(\omega)Q^{\sigma}_{\beta},
\end{equation}
with
\begin{eqnarray}
&& N_{\alpha\beta}(\omega) = \frac{k^2}{12}\sum_{\ell}
  \frac{\langle 1s|z_\alpha|d_{\ell}\rangle\langle d_{\ell}
       |z_\beta|1s\rangle}
       {\omega-(\epsilon_{\ell}-\epsilon_{1s})+i\Gamma}, \\
&&Q^\sigma= \left(\!\!\! \begin{array}{c}
              \sin\theta\sin\psi\\
             \frac{\sqrt{3}}{2}\cos\theta\sin{2\psi}\\
            -\frac{1}{\sqrt{2}}(\cos\theta\cos{2\psi}+\sin\theta\cos\psi)\\
             \frac{1}{\sqrt{2}}(\cos\theta\cos{2\psi}-\sin\theta\cos\psi)\\
             \frac 12 \cos\theta\sin{2\psi}
                  \end{array}\!\!\!\right), \\
&&Q^{\pi'}= \left(\!\!\! \begin{array}{c}
             \cos{2\theta}\cos\psi\\
             \frac{\sqrt{3}}{4}\sin{2\theta}(\cos{2\psi}-1)\\
             \frac{1}{2\sqrt{2}}(\sin{2\theta}\sin{2\psi}+2\cos{2\theta}\sin\psi)\\
             \frac{1}{2\sqrt{2}}(-\sin{2\theta}\sin{2\psi}+2\cos{2\theta}\sin\psi)\\
             \frac 14 \sin{2\theta}(\cos{2\psi}+3)
                  \end{array} \!\!\!\right). 
\end{eqnarray}
The incident and emitted photons have wave number
$k\approx 3.3\times 10^8$ cm$^{-1}$, since the K-edge absorption
energy is $6.552$ keV. 
The $\omega$ represents the photon energy.
The life-time broadening width $\Gamma$ of the $1s$ core hole
is assumed to be 1 eV. 
Energies $\epsilon_{1s}$, $\epsilon_j$ and $\epsilon_\ell$ correspond to 
the core state $|1s\rangle$ and the excited $p$ states, $|p_j\rangle$, and
$d$ state, $|d_\ell\rangle$, respectively. 
The dipole operators are defined as
$x_1=x$,
$x_2=y$ and
$x_3=z$,
and the quadrupole operators are
$z_1=(\sqrt{3}/2)(x^2-y^2)$,
$z_2=(1/2)(3z^2-r^2)$,
$z_3=\sqrt{3}yz$,
$z_4=\sqrt{3}zx$ and
$z_5=\sqrt{3}xy$.
We neglect the final-state interaction between the $1s$ core hole and
the excited electron. The excited $p$ electrons are so extended
that for the dipole process their effect is expected to be small,
but it may not be small for the quadrupole process due to the
localized nature of the $3d$ states.

Using the relations
$M_{11}^B(\omega)\simeq M_{22}^A(\omega)$,
$M_{22}^B(\omega)\simeq M_{11}^A(\omega)$, 
$M_{33}^B(\omega)\simeq M_{33}^A(\omega)$ and
$M_{12}^A(\omega)\simeq 0$,
we obtain for the dipole transition the AXS amplitude,
\begin{eqnarray}
&&J^A_{\sigma\to\pi'}(\omega)-J^B_{\sigma\to\pi'}(\omega) \cr
&&~~~~~~~~~~~~~=-\cos\theta\sin\psi(M_{11}^A(\omega)-M_{22}^A(\omega)).
\end{eqnarray}
For the quadrupole transition, we introduce the local coordinate frame 
according to
$x\to z$,
$y\to x$,
$z\to y$ at A sites, and
$x\to y$,
$y\to z$,
$z\to x$ at B sites.
Thereby the quadrupole operators are transformed such that
$z_1\to(-1/2)z_1\pm(\sqrt{3}/2)z_2$,
$z_2\to(\mp\sqrt{3}/2)z_1-(1/2)z_2$
(the upper and lower signs are for
A and B sites, respectively). 
Keeping only the diagonal terms 
$N_{11}^A(\omega)$ and $N_{22}^A(\omega)$ in these local coordinate frames,
we obtain for the quadrupole transition the AXS amplitude,
\begin{eqnarray}
&&L^A_{\sigma\to\pi'}(\omega)-L^B_{\sigma\to\pi'}(\omega)
=\frac 34 [\cos{2\theta}\cos\theta\cos\psi\sin{2\psi}+    \cr
&&\frac 12 \sin{2\theta}\sin\theta\sin\psi(\cos{2\psi}-1)] 
(N_{11}^A(\omega)-N_{22}^A(\omega)).
\end{eqnarray}
The $N_{11}^A(\omega)$ and $N_{22}^A(\omega)$ are proportional to
the $d_{y^2-z^2}$ and  $d_{3x^2-r^2}$ DOS's in the original coordinate frames.
The $d_{t_{2g}}$ DOS does not contribute to the AXS intensity
due to a cancellation between the amplitudes at A and B sites.

\begin{figure}
\begin{center}
\epsfile{file=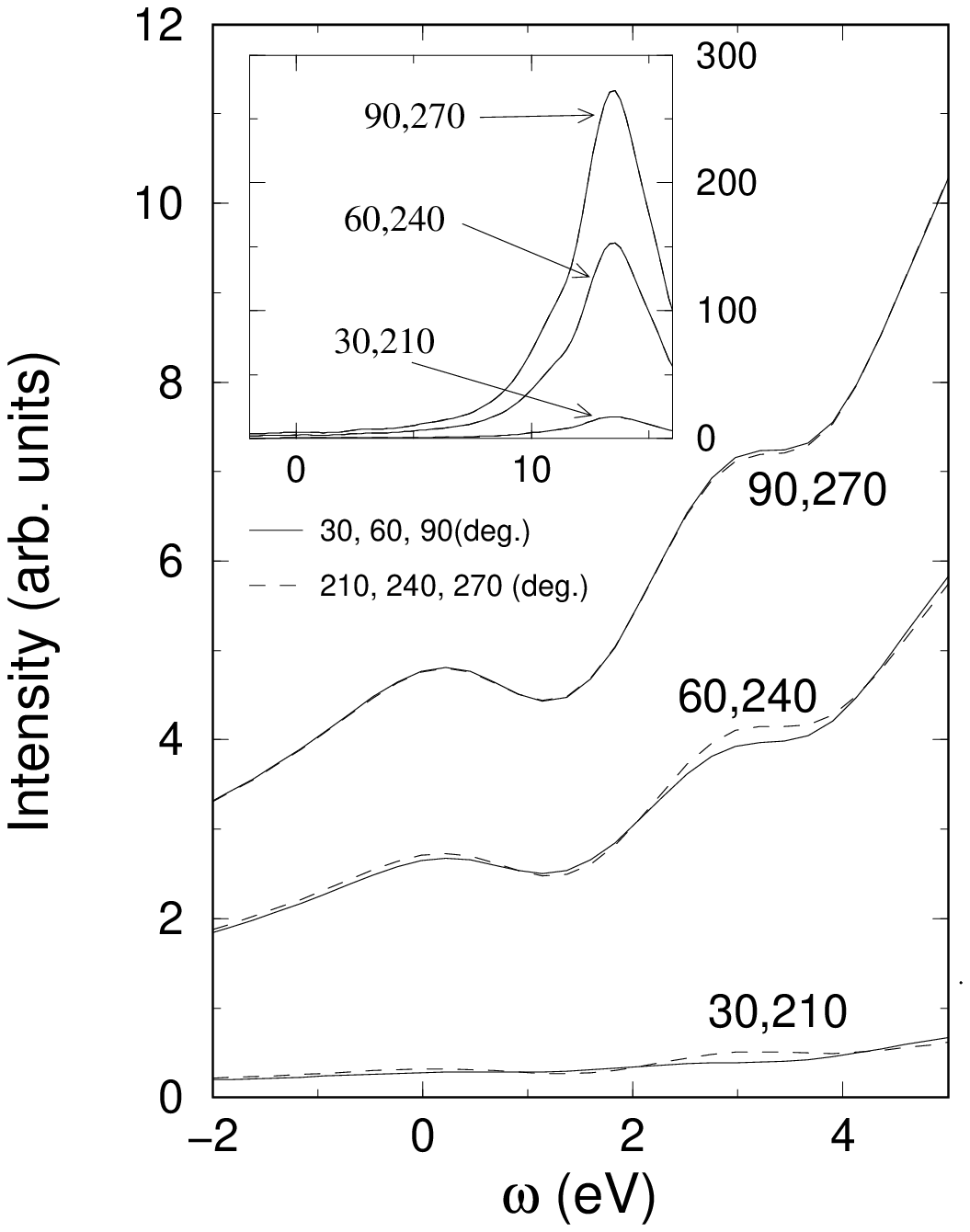,width=8cm}
\caption{
The AXS intensity of $\sigma\to\pi'$ scattering
associated with the $(1,0,0)$ reflection for several azimuthal angles,
as a function of the photon energy $\omega$.
The inset shows the overall spectra.
Here $\omega=0$ corresponds to the energy required for the transition
from the $1s$ state to the state at the top of the valence band.
The numbers attached to the curves represent the azimuthal angles.
}\label{fig.spec}
\end{center}
\end{figure}

Since the main peak of the $p$-symmetric DOS gives rise to 
the real part of the PKE amplitude,
the AXS intensity is enhanced by an interference between 
the tail of the main peak and the value brought about by the PKE DOS.
Figure \ref{fig.spec} shows the AXS intensity of $\sigma\to\pi'$
scattering corresponding to the $(1,0,0)$ reflection.
The PKE intensity is approximately $1/100$ of the main peak 
intensity, owing to the interference.
The PKE structure consists of two peaks; 
the one around $\omega\sim 0.5$ eV originates from the $p$-symmetric DOS 
of the majority spin states in region $a$, while 
the other around $\omega\sim 3$ eV originates from the $p$-symmetric DOS
of the minority spin states in regions $b$ and $c$ (see Fig.~\ref{fig.p}).
The fact that the solid curves are different from the broken ones
indicates an appreciable effect of the quadrupole transition.
 
\begin{figure}
\begin{center}
\epsfile{file=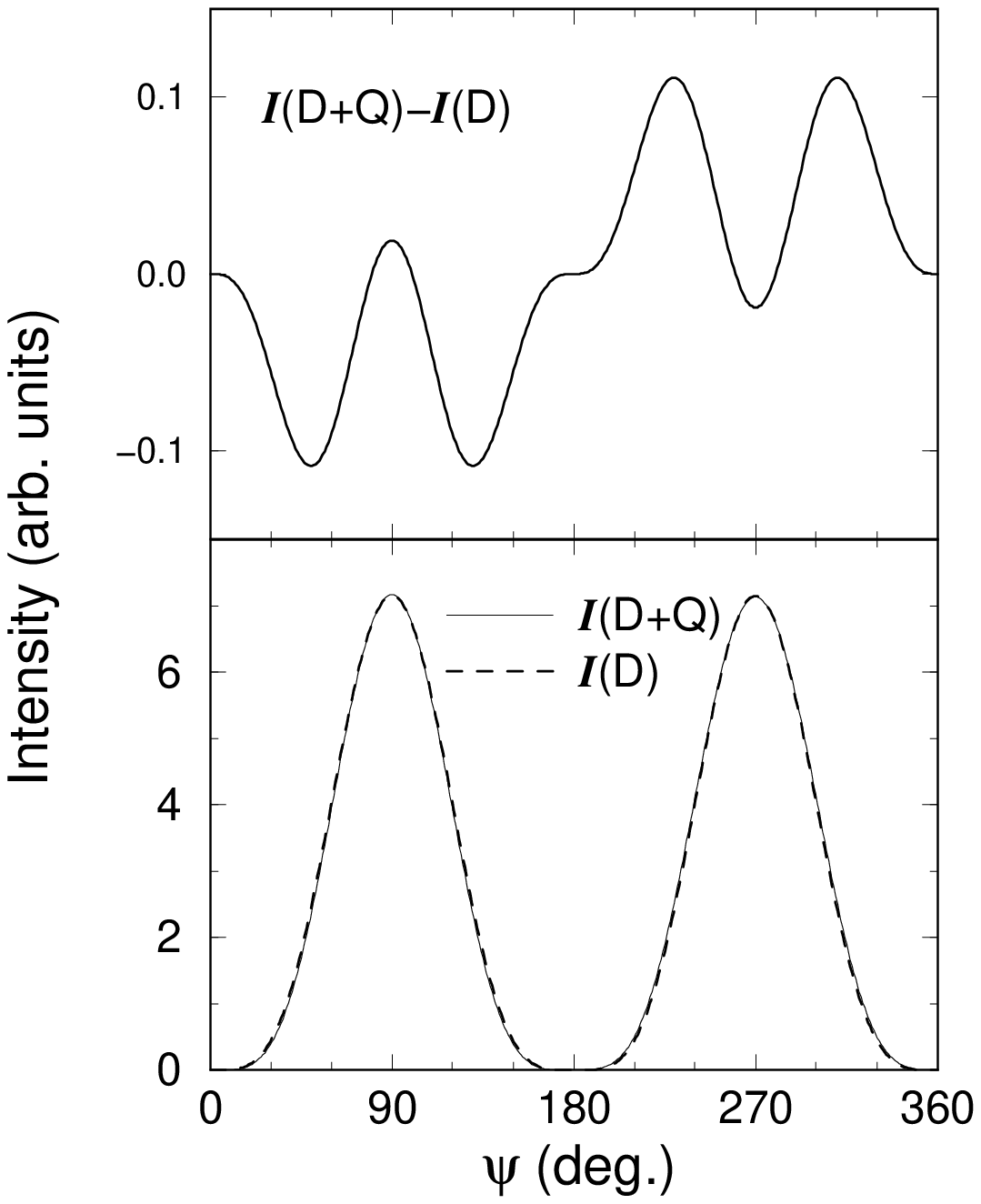,width=8cm}
\caption{
Azimuthal-angle dependence of the AXS intensity of the $\sigma\to\pi'$
transition at $\omega=3$ eV associated with the $(1,0,0)$ reflection.
The broken line represents the intensity coming from the dipole process.
The upper panel shows the difference of the intensities with and without
the quadrupole process. The scale of the ordinate is the same as
in Fig.~4.
}\label{fig.azimu}
\end{center}
\end{figure}

Figure \ref{fig.azimu} shows the azimuthal-angle dependence of the AXS
intensity at $\omega=3$ eV associated with the $(1,0,0)$ reflection.
As shown in the upper panel, the difference of the intensities with and
without the effect of the quadrupole transition for some angles
is negative. This is due to an interference between the
effects of the quadrupole process and the dipole one.
The effect of the quadrupole transition is one order of magnitude
smaller than that of the dipole transition.

In summary, we have calculated the anomalous X-ray scattering intensity,
using the KKR method based on the LDA.
We have found a two-peak structure in the pre-K-edge region
with an intensity approximately $1/100$ of that of the main 
peak. This originates from a mixing of $4p$ states of Mn to $3d$ states
of neighboring Mn sites. Therefore it is rather sensitive to the $3d$ states
and the orbital order, though the relation is not a direct one.
It seems dangerous to use a simple tight-binding model for
describing the $4p$ states, since they are so extended and 
the mixing is mediated by oxygen sites situated between Mn sites. 
The results of the present paper suggest that the JT distortion 
has a secondary effect on the PKE structure.
This contrasts with the main peak, which is generated by
the Jahn-Teller distortion.
\cite{Elfimov,Benfatto,Taka1}
The quadrupole effect shows a special azimuthal-angle dependence
through the interference with the dipole effect, 
but the effect is one order of magnitude smaller than the dipole effect.

One of the authors (J. I.) acknowledges a support from Japan Society 
of Promotion of Science. This work was partially supported by 
a Grant-in-Aid for Scientific Research from the Ministry of 
Education, Science, Sports and Culture.
%\newpage
\def\vol(#1,#2,#3){{\bf #1} (#2) #3}

\end{document}